\documentstyle{article}             %%
\def\bc{\begin{center}}           \def\ec{\end{center}}
\def\beq{\begin{equation}}         \def\eeq{\end{equation}}
\def\bear{\begin{eqnarray}}       \def\eear{\end{eqnarray}}
\def\bt{\begin{tabular}}          \def\et{\end{tabular}}
\def\la{\langle}                  \def\ra{\rangle}
\def\dg{\dagger}                  \def\ci{\cite}
\def\lb{\label}                   \def\ld{\ldots}
\def\lf{\left}                    \def\rt{\right}
                  
\def\pr{\prime}                   \def\sm{\small}

                   \def\pr{\prime}
                 \def\ts{\textstyle}
               \def\nn{\nonumber}

\def\Gam{\Gamma}     
\def\Dlt{\Delta}         
\def\lam{\lambda}    \def\Lam{\Lambda}    \def\sig{\sigma}

\begin{document}

\title{\bf State Extended Uncertainty Relations}
\author{D.A. Trifonov\\
        {\small Institute for Nuclear Research
        72 Tzarigradsko chauss\'ee, Sofia 1784, Bulgaria}}
\maketitle
\date {}

\begin{abstract}
A scheme for construction of uncertainty relations (UR) for $n$
observables and $m$ states is presented. Several lowest order UR 
are
displayed and briefly discussed. For two states $|\psi\ra$ and 
$|\phi\ra$
and canonical observables the (entangled) extension of Heisenberg 
UR reads
$[\Delta p(\psi)]^2[\Delta q(\phi)]^2+[\Delta p(\phi)]^2[\Delta q(\psi)]^2
\geq 1/2$.  Some possible applications of the new inequalities are 
noted.
\end{abstract}

%section 1
\section{Introduction}
The uncertainty principle is one of the basic principles in quantum
physics.
It was introduced by Heisenberg \ci{HK} on the example of the 
canonical
observables $p$ and $q$, and rigorously proved by Kennard \ci{HK} 
in the
form of the inequality $(\Dlt p)^2(\Dlt q)^2 \geq 1/4$, where $(\Dlt 
X)^2$
is the variance (dispersion) of $X$ (for the sake of brevity we work 
with
dimensionless observables).  This inequality is known as 
Heisenberg
uncertainty relation (UR) for $p$ and $q$.  It has been precised and
extended to arbitrary two quantum observables by Schr\"odinger 
and Robertson
\ci{SR} and to several observables by Robertson \ci{R}.

The Heisenberg UR became an irrevocable part of almost every 
textbook in
quantum physics while the interest in the more precise 
Schr\"odinger
\ci{SR} and Robertson \ci{R} UR has grown up only recently in 
connection
with the experimental generation of  squeezed states of the
electromagnetic field \ci{LK} and their generalization to arbitrary two
and several observables \ci{DKM,Wodki,T94,T97}.  Robertson UR 
has been
recently extended to all characteristic coefficients of the uncertainty
matrix \ci{TD}. Extensions of the Heisenberg UR to higher 
moments of $p$
and $q$ are made in \ci{SimMuk}.

The UR listed above, and perhaps all the other ones so far 
considered,
relate certain combinations of statistical moments of the 
observables
in one quantum state. The main aim of the present paper is to 
extend the
uncertainty principle to several states. The physical idea of such an
extension is simple: one can measure and compare the statistical 
moments of
two (or more) observables not only in one and the same state, but 
in two
(or more) different states. The Hilbert space model of quantum 
mechanics
permits us to derive easily such {\it state extended} UR.
The extended UR can be divided into two classes -- {\it entangled 
UR} and
nonentangled UR. An UR is called state entangled if it can not be
factorized over distinct states.

Next we first recall the ordinary characteristic UR which include the
known Schr\"odinger and Robertson ones (section 2) and then 
extend these UR to
several states (section 2). Some other state extended UR are also
established. Finally the simplest types of extended UR are 
displayed and
discussed briefly (section 3).  

\section{Characteristic UR}

The Schr\"odinger (or Schr\"odinger-Robertson)\ci{SR} UR for two
observables $X$ and $Y$ reads
\beq\lb{SUR}				%1
(\Dlt X)^2(\Dlt Y)^2-(\Dlt XY)^2 \,\geq\, \frac 14\lf|\la[X,Y]\ra\rt|^2,
\eeq
where $\la X\ra$ is the mean value of $X$ in a given state, $(\Dlt
X)^2 = \la X^2\ra - \la X\ra^2$ is the variance (the dispersion) of $X$,
and $\Dlt XY \equiv \la XY+YX\ra/2 -\la X\ra \la Y\ra$ is the 
covariance
of $X$ and $Y$. This UR was derived by Schr\"odinger from the 
Schwartz
inequality for the matrix element $\la\psi|(X-\la X\ra)(Y-\la 
Y\ra)|\psi\ra$.
The less precise inequality $(\Dlt X)^2(\Dlt Y)^2 \,\geq\, \frac
14\lf|\la[X,Y]\ra\rt|^2$ is usually called Heisenberg UR for $X$ and 
$Y$.

Robertson \ci{R} has formulated the uncertainty principle for several
observables $X_1,\ld,X_n$ in terms of an inequality between
determinants of the uncertainty matrix  $\sig(\vec{X};\psi)$
and the matrix $C(\vec{X};\psi)$ of the mean values of commutators 
of
$X_i$ and $X_j$,
\beq\lb{RUR}			   %2
\det \sig(\vec{X};\psi) - \det C(\vec{X};\psi) \,\geq\, 0,
\eeq
where $\vec{X} = X_1,\ld,X_n$,
$\sig_{ij} = (1/2)\la X_iX_j + X_jX_i\ra - \la X_i\ra\la X_j\ra$,
and  $C_{jk} = -(i/2)\la[X_j,X_k]\ra$.
For $n=2$ the inequality (\ref{RUR}) recovers (\ref{SUR}).  
Robertson has
first proved the nonnegative definiteness of the matrix 
$R(\vec{X};\psi)$
(to be called Robertson matrix), $R(\vec{X};\psi) =
\sig(\vec{X};\psi) + iC(\vec{X};\psi) \,\geq\, 0$.  This means that
all principle minors of $R$ are nonnegative.  For $n=2$ the 
inequality
(\ref{RUR}) coincides with $R(\vec{X};\psi)\geq0$.

Robertson UR hold for mixed states $\rho$ as well (see e.g.  the 
Appendix
in \ci{T99}). Recently \ci{TD} the UR (\ref{RUR}) has been
extended to all characteristic coefficients $C_r^{(n)}(\sig)$ of
$\sig(\vec{X};\rho)$ in the form
\beq\lb{CUR}			   %3
C^{(n)}_r(\sig) \,\geq\, C^{(n)}_r(C),\qquad r =
1,2,\ldots,n,
\eeq
where $C^{(n)}_r(C)$ is the characteristic coefficient of order $r$ of 
the
mean commutator matrix $C(\vec{X};\rho)$. The characteristic 
coefficients
of a matrix $M$ are defined by means of the secular equation 
\ci{Gant}
$\det(M - \lam) = \sum_{r=0}^{n} C^{(n)}_r(M)(-\lam)^{n-r}=0$. For 
$r=n$
one has $C^{(n)}_n(C) = \det C$, $C^{(n)}_n(\sig) = \det\sig$ and the
characteristic UR (\ref{CUR}) coincides with that of Robertson, eq.
(\ref{RUR}).

In comparison with the Heisenberg UR the Schr\"odinger and 
Robertson ones
have the advantage of being covariant under linear nondegenerate
transformations of the observables \ci{T97,TD}: if $\vec{X}\,^\pr =
\Lam\vec{X}$ then
\beq\lb{lt-obs}			   %4
\sig^\pr = \sig(\vec{X}\,^\pr;\rho) = \Lam\sig\Lam^T,\quad
C^\pr = C(\vec{X}\,^\pr;\rho) =  \Lam\,C\,\Lam^T.
\eeq
One sees that the equality in (\ref{RUR}) is invariant under
transformation (\ref{lt-obs}) with nonsingular $\Lam $, in particular 
with
$\Lam$ symplectic. For $r < n$ the equality in (\ref{CUR}) is 
invariant under (\ref{lt-obs}) if $\Lam$ satisfies $\Lam\Lam^T=1$.  If
$X_1,\ld,X_n$ close an algebra then Robertson UR is invariant under
algebra automorphisms.  In the case of orthogonal algebra the 
equality in
(\ref{CUR}) is invariant for every $r$.

The minimization of the  characteristic UR  and the relationship 
between
the minimizing states and the group-related coherent states and 
squeezed
states has been considered in \ci{T97,TD,T99}, the minimizing 
states being
called  characteristic optimal uncertainty states or characteristic
intelligent states of order $r$.
States which minimize Schr\"odinger UR (\ref{SUR}) were shown to 
be ideal
squeezed states for $X$ and $Y$ \ci{T94,T97}, while states which 
minimize
(\ref{RUR}) can be considered as squeezed states for several 
observables
\ci{T97}. 

\section{State extended UR}

It should be useful first  to recall that the derivation of the Robertson
UR resorts to the following\\

{\bf Lemma 1 (Robertson).} {\it If $H$ is a nonnegative definite 
Hermitian
matrix, then
\beq\lb{matRUR}			   %5
 \det S - \det A \geq 0,
\eeq
where $S$ and $A$ are the real and the imaginary part of $H$, 
$H=S+iA$}.

It is worth reminding that a matrix $H$ is nonnegative iff all its
principle minors $M_r(H)$ are nonnegative,
\beq\lb{H>0}			   %6
H\geq 0 \,\,\longleftrightarrow\,\, M_r(H) \geq 0,\quad r = 1,2,\ld,n.
\eeq
The proof of this lemma can be found in \ci{R}. With minor changes 
in the
notations it is reproduced in \ci{T99}. Robertson UR
(\ref{RUR}) corresponds to $H = R(\vec{X};\rho)$ in (\ref{matRUR}).
In \ci{TD} this lemma was extended to all principle minors and to all
characteristic coefficients of $H$,
\beq\lb{matCUR}			   %7
C^{(n)}_r(S) - C^{(n)}_r(A) \,\geq\, 0,\quad r = 1,2,\ldots,n.
\eeq
The characteristic UR (\ref{CUR}) correspond to 
$S=\sig(\vec{X};\rho)$ and
$A=C(\vec{X};\rho)$ in (\ref{matCUR}).

The state extension of the ordinary UR, which we shall derive 
below, are
based on the different physical choices of the matrix $H$ in 
(\ref{H>0}),
(\ref{matRUR}) and (\ref{matCUR}) and on the following lemma,\\

{\bf Lemma 2.} {\it If $H_\mu$  are nonnegative definite
Hermitian $n\times n$ matrices, $\mu=1,\ld,m$, then
\bear  		           %8   %9
C_r^{(n)}\lf(S_1+\ld +S_m\rt) \,-\,
 C_r^{(n)}\lf(A_1+\ld+A_m\rt) \,\geq\, 0, \lb{lemma2a}\\
C_r^{(n)}\lf(H_1+\ld+H_m\rt) \,-\,
C_r^{(n)}(H_1) - \ld - C_r^{(n)}(H_m)\,\geq\, 0,\lb{lemma2b}
\eear
where $S_\mu$ and $A_\mu$ are the real (and symmetric)
and the imaginary (and antisymmetric) parts of $H_\mu$}.

{\sl Proof}. The validity of (\ref{lemma2a}) immediately follows
from the Robertson lemma and its extension (\ref{matCUR}), and 
the known
fact that a sum of the Hermitian nonnegative matrices is a Hermitian
nonnegative matrix. We proceed with the proof of the inequality
(\ref{lemma2b}).  It is sufficient to establish it for $m=2$. Let $G$ 
and $H$
are Hermitian nonnegative definite matrices. We have to prove that
$C_r^{(n)}(G+H) - C_r^{(n)}(G) - C_r^{(n)}(H) \,\geq\, 0$. Since the
characteristic coefficients are sum of all principle minors \ci{Gant} it
is sufficient to consider the case of $r=n$, i.e. to prove the 
inequality
$\det(G+H) - \det G - \det H \,\geq\, 0$.\\
(a) Let one of the two matrices (say $G$) is positive definite.  Then 
both
$G$ and $H$ can be diagonalized by means of a unitary matrix $U$
\ci{Gant}, $G^\pr = U^\dg GU = {\rm diag}\{g_1,\ld,g_n\}$, $H^\pr = 
U^\dg
HU = {\rm diag}\{h_1,\ld,h_n\}$, and
\beq\lb{det(G+H)}		   %10
\det (G+H) =  {\ts\prod_i} (g_i+h_i) = {\ts\prod_i}\,g_i +
{\ts\prod_i}\, h_i + \Dlt,
\eeq
where $\Dlt = \det(G+H) - \det G - \det H= \det(G^\pr+H^\pr) - \det
G^\pr - \det H^\pr$, $i = 1,\ld,n$,
\beq\lb{Dlt}			   %11
\Dlt = g_1{\ts\prod_{j=2}^n}\, h_j +  g_1g_2{\ts\prod_{j=3}^n}\,h_j +
\ld +  h_1h_2{\ts\prod_{i=3}^n}\,g_i + h_1{\ts\prod_{i=2}^n}\,g_i.
\eeq
In view of $g_i >0$ and $h_j \geq 0$ all terms in (\ref{Dlt}) are
nonnegative, thereby $\Dlt \geq 0$. \\
(b) If both $G$ and $H$ are only nonnegative definite, then at
least one $g_i$ and one $h_j$ are vanishing, that is $\det G = 0
= \det H$ and from nonnegativity of the sum $G+H\geq 0$ and
(\ref{det(G+H)}) we obtain $\det(G+H) = \Dlt \geq 0$. End of the
proof.\\

{\bf Remark 1}. From the above proof it follows that if $\det\sum 
H_\mu =
\sum \det H_\mu$ then $\det H_\mu =0$, the inverse being not true.

Eqs. (\ref{lemma2a}) and (\ref{lemma2b}) can be called {\it extended
characteristic inequalities}. They are invariant under the similarity
transformation of the matrices $H_\mu$. At $m=1$ (one state) they 
 recover
the relations (\ref{matCUR}).\\

By a suitable physical choice of the nonnegative Hermitian matrices
$H_\mu$ in the inequalities (\ref{H>0}), (\ref{lemma2a}),
(\ref{lemma2b}) one can obtain a variety of UR for several states and
observables. We point out three physical choices of matrices 
$H_\mu$,
\beq\lb{H=R}               %12
H = R(\vec{X};\rho)= \sig(\vec{X};\rho) +
iC(\vec{X};\rho) \quad {\rm (Robertson\,\, matrix),}
\eeq
\beq\lb{H=G2}              %13
H  = \Gam(\chi_1,\ld,\chi_n)= R(\vec{X};\vec{\psi}),\qquad ||\chi_k\ra 
=
(X_k-\la\psi_k|X_k|\psi_k\ra)\,|\psi_k\ra ,
\eeq
\beq\lb{H=G3}              %14
H = \Gam(\phi_1,\ld,\phi_n) = G(\vec{X};\vec{\psi}),\qquad
||\phi_k\ra = X_k\,|\psi_k\ra,
\eeq
where $\Gam$ is the Gram matrix, $\Gam_{ij}(\Phi_1,\ld,\Phi_n) =
\la\Phi_i||\Phi_j\ra$, and  $|\psi_k\ra$ are normalized pure states.
The diagonal elements $R_{ii}(\vec{X};\vec{\psi})$ and
$R_{ii}(\vec{X};\rho)$  are just the variances of $X_i$ in the state
$|\psi_i\ra$ and (generally mixed) state $\rho$, while $G_{ii} =
\Gam_{ii}(\phi_1,\ld,\phi_n) = (\Dlt X_i(\psi_i))^2 +
\la\psi_i|X|\psi_i\ra^2$. Therefore the inequalities obtained in the 
above
scheme can be regarded as {\it state extended UR}. For brevity UR 
for $n$
observables and $m$ states should be called {\it UR of type (n,m)}.

For pure states
$|\psi_k\ra$ (\ref{H=R}) is a particular case of (\ref{H=G2}), and
the common structure of (\ref{H=R}) -- (\ref{H=G3}) is
$H = \Gam(\Phi_1,\ld,\Phi_n) = T(\vec{X},\vec{\psi}),$
where $\Phi_k$ denote the corresponding nonnormalized states
$|\Phi_k\ra$.
Let us note that $\Gam(\Phi_1,\Phi_2) \geq 0$ is equivalent to the
Schwartz inequality.
For one observable $X$ the matrix $G(X,\vec{\psi})$ is covariant 
under
linear transformation of states, 
$$|\psi^\pr_i\ra =
U^*_{ik}|\psi_k\ra \quad \rightarrow\quad G(\vec{X},\vec{\psi}\,^\pr) =
UG(\vec{X},\vec{\psi})U^\dg.$$ 
This property entails the invariance
of the equality in the extended highest order characteristic UR of 
type
(1,m), constructed by means of $G(X,\vec{\psi})$. If $UU^\dg=1$ 
then all
order extended characteristic UR of type (1,m) are invariant. 
Compare this
symmetry with that of ordinary characteristic UR under 
transformation
(\ref{lt-obs}). The extended UR of types (n,m) with $n,m > 1$ do not
possess such symmetry.

In all three cases of $H$ with pure states the UR $H\geq 0$ are
disentangled by means of linear transformations of states. The UR
corresponding to (\ref{lemma2b}) and (\ref{lemma2a}) are state 
entangled.
The proof of the nonentangled character of the UR $H\geq 0$ for
(\ref{H=R}), (\ref{H=G2}) and (\ref{H=G3}) with pure states can be 
easily
carried out using the diagonalization of $\Gam = 
\Gam(\Phi_1,\ld,\Phi_m)$.

\section{Extended UR of simplest types}

{\bf UR of type (1,2)}.
For $m=2$ (two states) the  choices $H=R(X;\vec{\psi})$ and
$H=G(X;\vec{\psi})$ in (\ref{H>0}), (\ref{lemma2a}) and 
(\ref{lemma2b})
produce two different UR which we write down as
\bear                   %16,  %17
(\Dlt X(\psi_1))^2\,(\Dlt X(\psi_2))^2 \,\geq\,
\lf|\la\psi_1|\,(X-\la\psi_1|X|\psi_1\ra)(X-\la\psi_2|X|\psi_2\ra)\,
|\psi_2\ra\rt|^2,\lb{(1,2)a}\\
((\Dlt X(\psi_1))^2+\la\psi_1|X|\psi_1\ra^2)((\Dlt X(\psi_2))^2+
\la\psi_2|X|\psi_2\ra^2) \,\geq\,
\lf|\la\psi_1|X^2|\psi_2\ra\rt|^2.\lb{(1,2)b}
\eear
Since the right hand sides of (\ref{(1,2)a}) and (\ref{(1,2)b}) are
generally greater than zero these inequalities reveal  correlations
between the statistical second moments of $X$ in different states.

These two UR are independent in the sense that none of them is 
more
precise than the other. To prove this suffice it to consider the 
example 
of $X=p$ and two Glauber coherent states. The minimization of
(\ref{(1,2)a}) and (\ref{(1,2)b}) occurs iff the two nonnormalized 
states
in the Gram matrix are proportional. In the case of (\ref{(1,2)a})
this is $(X-\la2|X|2\ra)|\psi_2\ra = \lam(X-\la1|X|1\ra) |\psi_1\ra$
wherefrom we easily deduce that if $X$ is a continuous observable
(such as $q$, $p$ or $p^2-q^2$ and $pq+qp$)
then UR (\ref{(1,2)a}) is minimized iff $|\psi_1\ra = |\psi_2\ra$.  It
follows from this condition that (\ref{(1,2)a}) and (\ref{(1,2)b}) can be
used for construction of distances between quantum states 
(observable
induced distances) \ci{T99}.\\

{\bf UR of type (2,1)}. For $n=2$ the inequalities (\ref{lemma2b}),
(\ref{lemma2a}) and (\ref{H>0}) coincide.  For two observables 
$X$,\,$Y$
and one state the Robertson choice (\ref{H=R}) coincides with 
(\ref{H=G2})
and when replaced in (\ref{H>0}) -- (\ref{lemma2b}) it produces 
Schr\"odinger
UR (\ref{SUR}). The choice (\ref{H=G3}) in (\ref{lemma2a}) and
(\ref{lemma2b}) generates the invariant UR
\beq\lb{(2,1)}             %18
[(\Dlt X)^2+\la X\ra^2)]\,[(\Dlt Y)^2+\la Y\ra^2)] \,\geq\,
  (\Dlt XY +\la X\ra\la Y\ra)^2  + \frac 14\lf|\la[X,Y]\ra\rt|^2,
\eeq
which however can be shown to be less precise than the 
Schr\"odinger one
(\ref{SUR}). The interpretation of any UR of type (2,1) is the same 
as
that of Schr\"odinger UR.\\

{\bf UR of type (2,2)}.
The number of possible UR of type (2,2) is much more.  The 
inequalities
$R(X,Y;\psi_1,\psi_2)\geq 0$ and $G(X,Y;\psi_1,\psi_2)\geq 0$ can 
be
displayed as ($\la i|X|i\ra \equiv \la\psi_i|X|\psi_i\ra$)
\bear                   %19,  %20
(\Dlt X(\psi_1))^2\,(\Dlt Y(\psi_2))^2 \,\geq\,
\lf|\la\psi_1|(X-\la1|X|1\ra)(Y-\la2|Y|2\ra)|\psi_2\ra\rt|^2,\lb{(2,2)a}
\\[3mm]
\,[(\Dlt X(\psi_1))^2+\la1|X|1\ra^2)]\,[(\Dlt Y(\psi_2))^2+\la2|Y|2\ra^2)]
\,\geq\, \lf|\la\psi_1|XY|\psi_2\ra\rt|^2. \lb{(2,2)b}
\eear
It is not difficult to establish (after some manipulations) that the
inequality (\ref{(2,2)b}) is less precise than (\ref{(2,2)a}). The
equalities in (\ref{(2,2)a}) and (\ref{(2,2)b}) are not
invariant under linear transformations of observables and/or states.  
The
equations (\ref{lemma2a}) and (\ref{lemma2b}) with $H_1 = 
R(X,Y;\psi_1)$ and
$H_2 = R(X,Y;\psi_2)$ both produce an entangled but very 
compact (2,2) UR,
\bear\lb{(2,2)Sa1}         %21
\frac 12\lf[(\Dlt X(\psi_1))^2(\Dlt Y(\psi_2))^2 + (\Dlt X(\psi_2))^2
(\Dlt Y(\psi_1))^2\rt] - \Dlt XY(\psi_1)\Dlt XY(\psi_2)\nn \\
\,\geq\,  \frac 14\la\psi_1|[X,Y]|\psi_1\ra\la\psi_2|[X,Y]|\psi_2\ra^*.
\eear
The equality in this relation is invariant under linear transformations 
of
$X,\,Y$, but not of $|\psi_1\ra$ and $|\psi_2\ra$.  With $|\psi_1\ra =
|\psi_2\ra$ in (\ref{(2,2)Sa1}) one recovers the Schr\"odinger UR
(\ref{SUR}). The inequality (\ref{(2,2)Sa1}) should be referred to as
{\it state extended Schr\"odinger UR}.  For the canonical $p$ and
$q$ it simplifies to
\beq\lb{(2,2)Spq}         %22
\frac 12\lf[(\Dlt p(\psi_1))^2(\Dlt q(\psi_2))^2 + (\Dlt p(\psi_2))^2
(\Dlt q(\psi_1))^2\rt] - \Dlt pq(\psi_1)\Dlt pq(\psi_2)
\,\geq\,  \frac 14.
\eeq
Similar to (but less precise than) (\ref{(2,2)Sa1}) is the (2,2) UR
obtained again from (\ref{lemma2a}) and (\ref{lemma2b}) with the 
third
choice (\ref{H=G3}).  The entangled UR (\ref{(2,2)Sa1}) admits less
precise version of the form  (corresponding to $\Dlt XY =0$) 
\beq\lb{(2,2)Hb}           %24
\frac 12\lf[(\Dlt X(\psi_1))^2\,(\Dlt Y(\psi_2))^2 +
(\Dlt X(\psi_2))^2\,(\Dlt Y(\psi_1))^2\rt] \,\geq\,
\frac 14\lf|\la\psi_1|[X,Y]|\psi_1\ra\,\la\psi_2|[X,Y]|\psi_2\ra\rt|.
\eeq
The latter inequality can be regarded as an entangled extensions of 
the
Heisenberg UR.  For $X=p$ and $Y=q$ the right hand side of 
(\ref{(2,2)Hb})
simplifies to $1/4$.

In view of the Remark 1 if UR (\ref{(2,2)Sa1}) is minimized then
(\ref{SUR}) is saturated by $|\psi_1\ra$ and $|\psi_2\ra$. Therefore
(\ref{(2,2)Sa1}) can be used for finer classification of Schr\"odinger
intelligent states.
From any extended UR one can obtain new ordinary UR by fixing 
all but one
of the states. For example if $|\psi_1\ra$ in (\ref{(2,2)Spq}) is fixed 
as a
canonical coherent state then (\ref{(2,2)Spq}) produces $(\Dlt p)^2 
+ (\Dlt
q)^2 \geq 1$. The latter UR is minimized in canonical coherent 
states only,
while the Heisenberg UR $(\Dlt p)^2(\Dlt q)^2\geq 1/4$ is minimized 
in any
squeezed state with $\Dlt pq = 0$.\\

{\bf UR of type (3,1) and type (2,m), $m > 2$}.
From higher order UR we note two cases: the case of
(3,1)-- UR corresponding to the choice $H_1 = G(X,Y;\psi)$,
$H_2(X,Z;\psi)$ in (\ref{lemma2b}), and the case of (2,m)-- UR
corresponding to the choice $H_\mu = R(X,Y;\rho_\mu)$ in the 
matrix
inequality (\ref{lemma2b}). After some consideration we arrive at the
explicite formulas for two entangled UR,
\beq\lb{(3,1)}             %
(\Dlt X)^2(\psi) [(\Dlt Y(\psi))^2 + (\Dlt Z(\psi))^2] \,\geq\,
2\Dlt XY(\psi) \Dlt XZ(\psi) + \frac 12 \la\psi|[X,Z]|\psi\ra
\la\psi|[Y,X]|\psi\ra,
\eeq

\bear\lb{(2,m)}             %
\sum_{\mu<\nu}^{m}\left[(\Dlt X(\rho_\mu))^2(\Dlt Y(\rho_\nu))^2 +
(\Dlt X(\rho_\nu))^2 (\Dlt X(\rho_\mu))^2\right]
- 2 \sum_{\mu<\nu}^{m}\Dlt XY(\rho_\mu) \Dlt YY(\rho_\nu) \nn \\
\,\geq\, 2\sum_{\mu<\nu}^{m}C(X,Y;\rho_\mu) C(X,Y;\rho_\nu),
\eear
where $C(X,Y;\rho) = -(i/2)\la[X,Y]\ra_{\rho} = (-i/2){\rm
Tr}(\rho[X,Y])$.\, With $Z=Y$ in (\ref{(3,1)}) one gets Schr\"odinger 
UR
(\ref{SUR}) for $X,\,Y$, while $X=Y$ produces new ordinary UR 
$(\Dlt
X(\psi))^2 + (\Dlt Z(\psi))^2 \geq 2\Dlt XZ(\psi)$.\,\,
At $m=2$ the inequality (\ref{(2,m)}) recovers (\ref{(2,2)Sa1}).

\section*{Conclusion}
We have established several classes of extended characteristic 
uncertainty
relations (UR) of type (n,m) for $n$ observables and $m$ states 
using the
Gram matrices of suitably constructed nonnormalized states.  
Entangled UR
can be obtained using characteristic inequalities
(\ref{lemma2b}) and (\ref{lemma2a}).

The extended UR reveal global statistical correlations of quantum
observables in distinct states. The characteristic inequalities could 
be
useful in many fields of mathematical and quantum physics, in 
particular
in the precise measurement theory. Extended UR can be
used for construction of observable induced distances between 
quantum
states and for finer classification of states, in
particular of group-related coherent states.

\end{document}